\documentclass[a4paper,10pt]{article}
\usepackage[utf8x]{inputenc}
\usepackage{amssymb}

\title{\bfseries\huge \vspace{-3.0cm} M-Theory   on  Complex  Spacetime }

\author{ \itshape by \\ \\ \bfseries Abdul Rouf Samurah \\\\ National Institute of Technology, Srinagar, Kashmir. \\\normalsize Email: abdulrouf.lightspeed@yahoo.com}

\begin{document}

\maketitle

\begin{abstract}
In this paper we will analyse   ABJM theory in
 $N=1$ superspace formalism on complex   spacetime. 
We will then analyse the BRST 
and anti-BRST symmetries for  this   
  theory.
  We will show that 
the sum of gauge fixing 
and  ghost terms  for this   theory can be expressed 
as a combination of  the total BRST or the total anti-BRST variations. 

\end{abstract}

\section{Introduction}
Gauge symmetry plays an important role in nature \cite{c}-\cite{d}.
It is also important in understanding 
  multiple M2 branes \cite{a1}-\cite{a2}, and 
 multiple M5 branes \cite{b1}-\cite{b2}. 
Gauge symmetry can be anaysed either via the Wheeler-DeWitt approach 
\cite{1}-\cite{2}
or the BRST approach \cite{1a}-\cite{2a}.
Recently interest in complex spacetime has  been generated due to 
certain developments in the string 
theory \cite{1b}-\cite{4sy}. 
Complex spacetime has been  studied as a model for non-symmetric
 gravity formalism \cite{1sy,2sy}.
 Nonanticommutative field theories and 
nonanticommutative quantum gravity have  been formulated 
 on this complex spacetime \cite{sys2}-\cite{sys4}.

The action for $M$-theory  at low energies  is a superconformal 
action with 
with manifest $N =8$ supersymmetry. This action was discovered
by  Bagger and Lambert  and was based on 
gauge symmetries generated by a Lie $3$-algebra 
\cite{1}-\cite{5}.  However, only one example of 
such a such $3$-algebra is known  and  so far the   rank of the gauge 
group has not been increased. 
So, a  $U(N)_k \times U(N)_{-k}$ superconformal Chern-Simons-matter 
 theory with level $k$ and $-k$ with arbitrary rank and   
$N = 6$ supersymmetry was also  constructed \cite{apjm}. 
This theory called the ABJM theory. It  is also thought to describe the 
the action of  $M$-theory as it reduced to the Bagger-Lambert theory
for the only known example of Lie $3$-algebra. Its   
 supersymmetry can also get enhanced to $N = 8$ supersymmetry \cite{su}.
 Furthermore,  a $SO(8)$ $R$-symmetry at Chern-Simons levels
$k = 1,2$ also exists for this model. 

BRST and anti-BRST symmetries
 for the ABJM theory has been studied \cite{brsta, zcsddc9}. This
symmetry for the in deformed superspace. In this 
paper we will study the BRST and anti-BRST symmetries
 for the ABJM theory on deformed complex spacetime. 
It will be demonstrated that on complex spacetime, 
just like in the the deformed superspace, the 
sum of the gauge fixing and ghost terms can be expressed 
as a total BRST and anti-BRST variation.

\section{Deformation of ABJM Theory }
In this section we will deform the superspace of ABJM theory on
complex spacetime. 
We will now impose the following non-anticommutativity 
relation \cite{sys4} 
\begin{equation}
 \{ x^\mu, x^\nu\} = 2  x^\mu x^\nu + i \tau^{\mu\nu},
\end{equation}
 where $\tau^{\mu\nu} = \tau^{\nu\mu}$. 
 This leads to the following star product 
\begin{equation}
 X (x, \theta )\Diamond Y (x, \theta)= \exp \left[\frac{i  }{2}\tau^{\mu\nu}\partial_\mu^1
\partial_\nu^2 \right] X (x_1, \theta ) Y (x_2, \theta)_{
x_1 = x_2 =x}.
\end{equation} 
It is also useful to define the following bracket 
\begin{equation}
 2[X, Z]_{\Diamond} =  X \Diamond Z \pm  Z \Diamond X, 
\end{equation}
where the relative sign is negative unless  both the fields are fermionic. 
The super-derivative $D_a$ is defined by 
\begin{equation}
 D_a = \partial_a + (\gamma^\mu \partial_\mu)^b_a \theta_b,
\end{equation}
Now we define  the spinor superfieds 
$\Gamma_a$ and $\tilde \Gamma_a$ as
\begin{eqnarray}
 \Gamma_a = \chi_a + B \theta_a + \frac{1}{2}(\gamma^\mu)_a A_\mu + i\theta^2 \left[\lambda_a -
 \frac{1}{2}(\gamma^\mu \partial_\mu \chi)_a\right], \nonumber \\
 \tilde\Gamma_a = \tilde\chi_a + \tilde B \theta_a + \frac{1}{2}(\gamma^\mu)_a \tilde A_\mu + i\theta^2 \left[\tilde \lambda_a -
 \frac{1}{2}(\gamma^\mu \partial_\mu \tilde\chi)_a\right]. 
\end{eqnarray}
We also define scalar superfieds $X^I$ and $X^{I \dagger}$, with
\begin{eqnarray}
 \nabla_{(X)}^a \Diamond X^{I } &=& D_a  X^{I } + i \Gamma_a \Diamond X^I - i \tilde\Gamma_a \Diamond X^I, \nonumber \\ 
 \nabla_{(X)}^a \Diamond X^{I \dagger} &=& D_a  X^{I  \dagger} 
- i \Gamma_a \Diamond X^{I  \dagger}  + i \tilde\Gamma_a \Diamond X^{I  \dagger},
\end{eqnarray}
Now we the ABJM theory given by
\begin{equation}
{ \mathcal{L}_c} =  \mathcal{L}_{M} + \mathcal{L}_{CS} 
- \tilde{\mathcal{L}}_{CS},
\end{equation}
where
\begin{eqnarray}
 \mathcal{L}_{CS} &=& \frac{k}{2\pi} \int d^2 \,  \theta \, \,   Tr [  \Gamma^a  \Diamond  \omega_a] _|, 
\nonumber \\
 \tilde{\mathcal{L}}_{CS} &=& \frac{k}{2\pi} \int d^2 \, 
 \theta \, \,   Tr [  \tilde{\Gamma}^a  \Diamond  \tilde{\omega}_a] _|, 
\end{eqnarray}
 and
\begin{eqnarray}
 \omega_a &=&  \frac{1}{2} D^b D_a \Gamma_b - i  [\Gamma^b, D_b \Gamma_a]_{\Diamond } - \frac{1}{3} [ \Gamma^b ,
[ \Gamma_b, \Gamma_a]_{\Diamond}]_{\Diamond}\nonumber \\
   \tilde\omega_a &=&  \frac{1}{2} D^b D_a \tilde\Gamma_b -i  [\tilde\Gamma^b, D_b \tilde\Gamma_a]_{\Diamond } - 
\frac{1}{3} [ \tilde\Gamma^b,
[ \tilde\Gamma_b,  \tilde\Gamma_a]_{\Diamond } ]_{\Diamond }. 
\end{eqnarray}
Furthermore, 
\begin{eqnarray}
 \mathcal{L}_{M} = \frac{1}{4} \int d^2 \,  \theta \, \,  Tr \left[ \nabla^a_{(X)} \Diamond  X^{I \dagger}  \Diamond \nabla_{a (X)} \Diamond  X_I ] +
\frac{4\pi}{k} \mathcal{V}[X^{I\dagger}, X^I]_{ \Diamond }\right]_|,
\end{eqnarray}
where $\mathcal{V}[X^{I\dagger}, X^I]_{ \Diamond }$ 
is the potential term  with the 
product of all fields replaced by the star product. 

\section{BRST and anti-BRST Symmetry}
Some of  the degree's of freedom in the Lagrangian 
 are not
 physical. This is because  of the following gauge transformations, 
\begin{eqnarray}
 \delta\, \Gamma_a = \nabla_a \Diamond \Lambda, &&  
 \delta\, \tilde\Gamma_a = \tilde\nabla_a \Diamond \tilde\Lambda, \nonumber \\ 
\delta\, X^{I } = 
i(\Lambda \Diamond X^{I } - X^{I } \Diamond\tilde \Lambda
 ),  &&  \delta \, X^{I \dagger  } = i(\tilde \Lambda\Diamond  
X^{I\dagger  } -   
X^{I\dagger  }\Diamond  \Lambda),
\end{eqnarray}
here 
\begin{eqnarray}
 \nabla_a = D_a -i \Gamma_a, && \tilde{\nabla}_a  = D_a - i \tilde{\Gamma}_a. 
\end{eqnarray}
So, we add the following  gauge fixing term to 
the original Lagrangian density,
\begin{equation}
\mathcal{L}_{gf} = \int d^2 \,  \theta \, \, Tr  \left[b \Diamond (D^a \Gamma_a) + \frac{\alpha}{2}b \Diamond b -
i\tilde{b}  \Diamond (D^a \tilde{\Gamma}_a) + \frac{\alpha}{2}\tilde{b}  \Diamond \tilde b 
\right]_|. 
\end{equation}
We also add the following 
 ghost term  
\begin{equation}
\mathcal{L}_{gh} = \int d^2 \,  \theta \, \,  Tr 
[ \overline{c}  \Diamond D^a \nabla_a  \Diamond c - \tilde{\overline{c}}  \Diamond D^a \tilde{\nabla}_a \Diamond \tilde{c} ]_|.
\end{equation}
The total Lagrangian density 
obtained this way is invariant under the
 following BRST transformations, 
\begin{eqnarray}
s \,\Gamma_{a} = \nabla_a \Diamond  c, && s\, \tilde\Gamma_{a} =\tilde\nabla_a  \Diamond  \tilde c, \nonumber \\
s \,c = - {[c,c]}_ {\Diamond} , && s \,\tilde{\overline{c}} =- \tilde b - 2 [\tilde{\overline{c}} ,  \tilde c]_{\Diamond}, \nonumber \\
s \,\overline{c} = b, && s \,\tilde c = - [\tilde c, \tilde c]_{\Diamond}, \nonumber \\ 
s \,b =0, &&s \, \tilde b= - [ \tilde b, \tilde{\overline{c}}]_{\Diamond}, \nonumber \\ 
s \, X^{I } = i(c \Diamond X^{I } -  X^{I }\Diamond \tilde c), 
 &&  s \, X^{I \dagger } = i(\tilde c\Diamond  X^{I \dagger } 
- X^{I \dagger }\Diamond   c),
\end{eqnarray}
and the following BRST transformations, 
\begin{eqnarray}
\overline{s} \,\Gamma_{a} = \nabla_a  \Diamond \overline{c}, &&  \overline{s} 
\, \tilde \Gamma_{a} =  \tilde\nabla_a \Diamond \tilde{\overline{c}},\nonumber \\
\overline{s} \,c = -b - 2 [\overline{c}, c]_ \Diamond,  && [\overline{s} ,\tilde c]_ \Diamond =  \tilde b, \nonumber \\
\overline{s} \,\overline{c} = - [\overline{c}, \overline{c}]_\Diamond, 
                   &&\overline{s} \,\tilde{\overline{c}} = -[ \tilde{\overline{c}},\tilde{\overline{c}}]_\Diamond,\nonumber \\ 
\overline{s} \,b =- {[b,c]}_\Diamond,  && \overline{s}  \,\tilde b  = 0, \nonumber \\ 
\overline{s} \, X^{I } = i(\overline{c} \Diamond X^{I } -  X^{I }\Diamond
 \tilde{\overline{c}}), &&   \overline{s} \, X^{I \dagger } = 
i(\tilde{\overline{c}}\Diamond 
 X^{I \dagger } - X^{I \dagger }\Diamond  \tilde{\overline{c}}).
\end{eqnarray}
Both these sets of transformations are nilpotent. 
\begin{equation}
[s, s]_\Diamond = [\overline{s}, \overline{s}]_\Diamond  = 0. 
\end{equation}
We can now write 
\begin{eqnarray}
\mathcal{L}_{gf} + \mathcal{L}_{gh} 
 &=&- \int d^2 \,  \theta \, \, \overline{s}\, Tr  \left[ c \Diamond \left(D^a \Gamma_a   
 -  \frac{i\alpha}{2}b\right) - \tilde c \Diamond \left(D^a \tilde\Gamma_a   
 -  \frac{i\alpha}{2}\tilde b\right) \right]_|\nonumber\\
 &=&  \int d^2 \,  \theta \, \,  s\, Tr \left[ \overline{c} \Diamond \left(D^a  \Gamma_a 
 -  \frac{ \alpha}{2}b\right)
 -   \tilde{\overline{c}} \Diamond \left(D^a  \tilde \Gamma_a  
 -  \frac{ \alpha}{2}\tilde b\right)
\right]_|.
\end{eqnarray}
Thus, the sum of gauge fixing  and ghost terms 
can be expressed either as a total BRST variation
or as a total anti-BRST variation.

\section{Conclusion}
In this paper we studied the 
 ABJM theory in $N=1$ superspace formalism. 
We have then studied the deformation of this theory on deformed 
complex spacetime. 
We 
 analysed  the BRST 
and the anti-BRST symmetries
 for  this     theory. It was shown that 
the sum of the ghost and gauge fixing terms can 
be written as a total BRST or a total anti-BRST variation. 
It will be interesting to perform this analyses with 
non-linear BRST and anti-BRST transformations. 
It is known that for ABJM theory 
on a deformed superspace, in  Landau and Non-linear gauges, the 
sum of the gauge fixing and ghost terms can be expressed 
as a combination of both BRST and anti-BRST transformations. 
It will be interesting to derive a similar result for 
the ABJM theory on complex spacetime.

\end{document}